\RequirePackage[2020-02-02]{latexrelease}
\documentclass[aps,twocolumn,
10pt,aps,amsmath,amssymb,nofootinbib,superscriptaddress]{revtex4}
\usepackage{braket}
\usepackage{sans}
\usepackage{hyperref}
\usepackage{newtxtext,newtxmath}
\usepackage{tikz}

\usepackage{booktabs,multirow,tabularx}
\usepackage{blindtext}
\usepackage{float}
\usepackage{physics}
\usepackage{pbox}
\usepackage{url}
\usepackage{eqnarray,amsmath}
\usepackage{empheq}
\usepackage[export]{adjustbox}
\usepackage{graphicx}
\usepackage{makecell}
\usepackage{bbold}
\usepackage{ amssymb }
\begin{document}
\title{Probing non-unitarity of neutrino mixing in the scenario of Lorentz violation and dark nonstandard interaction}
 	\author{Trisha Sarkar}
	\email{sarkar.2@iitj.ac.in}
	\affiliation{Indian Institute of Technology Jodhpur, Jodhpur 342037, India}
	
 \begin{abstract}
Neutrino flavour oscillation is one of the primary indication of the existence of physics beyond standard model. The presence of small neutrino mass is indispensable to explicate the oscillation among different flavours of neutrino. By the addition of a right handed neutral lepton with the standard model fermions, it is possible to generate tiny neutrino mass. Such additional fermion may induce non-unitarity to the $3\times 3$ PMNS mixing matrix which influences the propagation of neutrino in space-time. In this work the effect of both unitary and non-unitary mixing matrix is analyzed in neutrino oscillation in presence of two new physics scenarios, Lorentz violation and dark non-standard interaction. Lorentz symmetry violation mainly appears at the Planck scale, which may also be manifested at a lower energy level. On the other hand, dark non standard interaction arises due to the interaction of neutrino with the environmental dark matter which contributes as a perturbative correction to the neutrino mass. The analysis is carried out in the context of long baseline DUNE and short baseline Daya Bay experimental set up. The signature of dark nonstandard interaction is observable in both DUNE and Daya Bay set up in terms of large value of neutrino survival and oscillation probability respectively and is a possible explanation for the excess flux observed at $\sim5$ MeV in Daya Bay experiment. The signature of Lorentz violation is also possible to be observed in the Daya Bay experiment.
\end{abstract}
\maketitle
\section{Introduction}\label{intro}
Neutrino oscillation is a well established phenomenon at present in the domain of particle physics. The presence of small neutrino mass is essentially required for the neutrino flavour oscillation to occur which is one of the most primary necessity to look for the physics beyond standard model (SM). While in SM neutrinos are considered to be left handed (LH) massless Weyl fermions, in several extensions of SM neutrino mass may  appear due to some radiative correction \cite{Zee:1980ai,Frigerio:2002in,Ahriche:2015loa} or by the inclusion of right handed (RH) chiral neutrinos \cite{Kersten:2007vk,Tang:2014hna,King:2003jb}.
In generally, the non zero neutrino mass is expected to appear due to breaking of some hidden symmetry \cite{MohseniSadjadi:2017jne,Lindner:2014oea}. The smallness of neutrino mass can be explained from type-I Seesaw mechanism which involves the addition of heavy neutral leptons \cite{Branco:2020yvs}. 

Due to the presence of these additional leptons the unitary neutrino mixing matrix, which is a subset of a general lepton mixing matrix, acquires non-unitary properties. 
Such non-unitarity condition affects the neutrino oscillation and its propagation through the medium. Non unitraity (NU) of mixing matrix also gives rise to lepton flavour violation (LFV) which is another signature of beyond standard model (BSM) phenomena \cite{Forero:2011pc}. The Seesaw scale may exist at any point within electroweak (EW) and Planck scale. Although high scale Seesaw mechanism are unable to provide momentous deviation from unitarity, several low scale mechanisms are capable of doing so \cite{Malinsky:2009df, Forero:2021azc}. The effect of non-unitary mixing matrix has been discussed in several previous works in context of different aspects of neutrino oscillation \cite{Berryman:2014yoa, Cai:2017mow, Miranda:2019ynh}. In addition to the feature of non-unitarity of the mixing matrix, two of the other most profound BSM scenarios in the neutrino sector are the violation of Lorentz symmetry and the presence of nonstandard interaction (NSI). 

Lorentz symmetry is conserved in two of the foremost theories of nature, SM of particle physics and general theory of relativity (GR). Various models have been developed allowing the spontaneous violation of Lorentz symmetry at the Planck scale ($\sim 10^{19}$ GeV) which unifies SM with gravity. Although it is impossible to test any theory at this energy scale, it is expected that such new physics (NP) signals are manifested at lower energy scale which can be probed with current experimental facilities. Mathematically, SM is portrayed as the renormalizable effective field theory (EFT) at lower energy scale consisted of the operators with dimension $D\leq 4$, while the NP signals are represented in terms of non-renormalizable higher dimensional operators ($D>4$) suppressed by the scale of NP. The technique of interferometry has been widely utilized to investigate the Lorentz symmetry \cite{Michelson:1887zz, PhysRev.42.400, Kobakhidze:2007iz, 2017arXiv170709016P}. Neutrino oscillation provides a natural source of interferometer \cite{CaboMontesdeOca:2020xeq} which is an excellent tool to search for Lorentz violation (LV). Currently ongoing oscillation experiments are able to search for the violation of such a fundamental symmetry. Several experimental facilities have constrained the strength of LV \cite{Barenboim:2018ctx,Super-Kamiokande:2014exs,Glashow:2004im}.

Non standard interaction is another elegant way to incorporate NP in the neutrino sector. Introduced by Wolfenstein in 1978 in terms of flavour changing neutral current (FCNC) when the neutrino undergoes the coherent forward elastic scattering in the matter \cite{PhysRevD.17.2369}, NSI has been widely studied in many literatures \cite{Farzan:2017xzy, Proceedings:2019qno,Ohlsson:2012kf,Antusch:2008tz}. It should be mentioned that the discussions in these works involve neutrino interaction mediated by a vector boson, which directly contributes to the matter potential, while the neutrinos may also interact via a scalar mediator, giving rise to scalar NSI \cite{Ge:2018uhz,Medhi:2021wxj}. Since the universe contains a large abundance of dark matter (DM), it is also possible that the scalar particle may belong to the dark sector, generating dark NSI \cite{article}. Unlike the case of vector NSI, which is possible to be represented in terms of dimension-$6$ four fermion operator and introduces modification to the matter potential, the neutrino-scalar interaction induces correction to the neutrino mass term. The bounds on the NSI parameters corresponding to the scalar mediator is much weaker, as compared to the case of vector NSI. Borexino collaboration has been able to put constraints on the scalar NSI parameters \cite{Borexino:2017rsf, Khan:2019jvr}.

In this work, the comparative analysis of the unitary and non-unitary mixing matrix are presented, in the sector of Lorentz violation (LV) and dark NSI within the framework of neutrino oscillation. The results are determined in terms of neutrino survival and oscillation probability in the context of long baseline (LBL) accelerator experiment, DUNE and short baseline (SBL) reactor experiment, Daya Bay. The signature of dark NSI with the consideration of  NU scenario is observable in DUNE experiment within the energy range $\sim 2.5-2.8$ GeV at which the maximum flux is obtained. It is also identified that the same combined scenario of dark NSI with NU mixing could be a possible explanation for the excess flux observed at $\sim 5$ MeV in Daya Bay, known as the "$5$ MeV bump" \cite{Jaffe:2021wgf,Dentler:2017tkw,Berryman:2018jxt}.

The plan of this work is as follows. In section \ref{form} the formalism of this work is described including the different aspects of NP scenarios. In section \ref{result}, the results of this analysis are discussed and finally the conclusion is mention in section \ref{conclusion}.

\section{Formalism}\label{form}
In this section the formalism of this work is presented. In section \ref{NU}, the facet of non-unitarity of neutrino mixing matrix is discussed, while the scenarios of LV and dark NSI are described in section \ref{LV} and \ref{NSI} respectively. In section \ref{matter}, the analytical framework to determine the neutrino oscillation probability is presented.

\subsection{Non unitarity of mixing matrix}\label{NU}
The origin of non-unitary neutrino mixing lies in the existence of non zero neutrino mass. Recent measurement by KATRIN collaboration estimates the neutrino mass to be $m_{\nu}<0.8$ eV \cite{KATRIN:2022ayy}. Neutrino mass can be either of Dirac or Majorana in nature. However, it is not favourable for the neutrino to acquire pure Dirac mass term, as the corresponding Yukawa couplings are extremely small $\sim 10^{-12}$  for the generation of neutrino mass $m_{\nu}\sim 0.1$ eV \cite{deGouvea:2016qpx} which is much smaller as compared to the other SM fermions. This also demands the lepton number to be a fundamental symmetry of SM, which originally appears accidentally in SM. Such problem is possible to overcome if the neutrinos are considered to be of Majorana nature which beings out the concept of the well known Seesaw mechanism \cite{Lindner:2001hr}. This can be achieved by the addition of non-renormalizable  and lepton number violating dimension-$5$ Weinberg operator with the SM effective Lagrangian, which generates tiny Majorana neutrino mass after electroweak symmetry breaking (EWSB). 
If the mass of the RH neutrino is much larger than the vacuum expectation value of the SM Higgs boson ($\sim 246$ GeV), it is possible to generate tiny neutrino mass, without undergoing the problem of small Yukawa coupling \cite{Kanemura:2011jj}.
 
However, the addition of the RH neutrinos affects the flavour oscillation and the interaction nature of the neutrinos. While incorporating the RH neutrino with its LH counterpart, the contribution of RH neutrino to the weak CC reaction gives rise to NU property of the $3\times 3$ PMNS mixing matrix. In the presence of $n$ number of neutrinos, the generic mixing matrix takes the following form
\begin{equation}\label{eq1}
U^{n\times n}=\begin{pmatrix}
T^{3\times 3} & Q^{3\times (n-3)} \\
V^{(n-3)\times 3} & U^{(n-3)\times(n-3)},
\end{pmatrix}
\end{equation}
where $T^{3\times 3}$ and $U^{(n-3)\times (n-3)}$ represent the mixing in the sector of LH and RH neutrinos respectively, provided there are three active LH neutrinos and $(n-3)$ number of RH neutrinos. $ Q^{3\times (n-3)}$ and $V^{(n-3)\times 3}$ contain the coupling parameters in the RH sector. In case of unitary mixing, $T^{3\times 3}$ is represented as the PMNS mixing matrix, given by
 \begin{align}\label{m}
& U_{PMNS}= \nonumber \\ 
& \begin{pmatrix}
c_{12}c_{13} & s_{12}c_{13} & s_{13}e^{-i\delta} \\
-s_{12}c_{23}-c_{12}s_{13}s_{23}e^{i\delta} & c_{12}c_{23}-s_{12}s_{13}s_{23}e^{i\delta} & c_{13}s_{23} \\
s_{12}s_{23}-c_{12}s_{13}c_{23}e^{i\delta} & -c_{12}s_{23}-s_{12}s_{13}c_{23}e^{i\delta} & c_{13}c_{23} \end{pmatrix}.
\end{align}
Here $c_{ij}=cos \theta_{ij}$, $s_{ij}=sin \theta_{ij}$ and $\delta$ is the CP violating phase. In presence of NU,
$T^{3\times 3}$ can be parametrized in a model independent way, assuming that the additional neutral heavy leptons do not contribute to neutrino oscillation \cite{Forero:2021azc}. One of the most convenient parametrization of $T^{3\times 3}$ is expressed as follows \cite{Fernandez-Martinez:2007iaa, Escrihuela:2015wra}
\begin{equation} \label{eq2}
T^{3\times 3}=\begin{pmatrix}
\alpha_{00} & 0 & 0 \\
\alpha_{10} & \alpha_{11} & 0 \\
\alpha_{20} & \alpha_{21} & \alpha_{22} 
\end{pmatrix} U_{PMNS},
\end{equation}
where $\alpha_{ij}$'s are the parameters that invoke NU and $U_{PMNS}$ is given by eqn. \eqref{m}. In generally, the diagonal parameters are real , while the off-diagonal ones are complex ($\alpha_{ij}=|\alpha_{ij}|e^{\phi_{ij}},~i \neq j$) that can provide an additional source of CP violation. 
The constraints on NU parameters in eqn. \eqref{eq2} are obtained from several decay processes \cite{Antusch:2006vwa,Fernandez-Martinez:2007iaa,Escrihuela:2019mot} and oscillation experiments \cite{Escrihuela:2015wra,Miranda:2019ynh}. 

\subsection{Violation of Lorentz symmetry}\label{LV}
It is widely known that SM is invariant under Lorentz transformation, which leads to the conservation of CPT symmetry in the universe. It is assumed in several extensions of SM that the Lorentz symmetry is possible to be violated at some higher energy scale, possibly at Planck scale. The origin of Lorentz violation (LV) is discussed extensively in various theoretical models of string theory \cite{Mavromatos:2007xe} and quantum gravity \cite{Cuadros-Melgar:2011whf, Moffat:2009ks,Eichhorn:2019ybe}. In string theory, the violation of Lorentz symmetry can occur spontaneously in presence of a non-perturbative vacuum generating tensor fields acquiring non-zero vacuum expectation value \cite{Bluhm:2013mu,2014IJMPS..3060274H}. In non-commutating field theory, the phenomenon of Lorentz violation occurs naturally. In quantum gravity theory, the Lorentz violation may appear by introducing the concept of anisotropic scaling of the space and time \cite{Horava:2009if}.

The observation of Lorentz violation is possible at Planck scale ($E_{Pl}\sim 10^{19}$ GeV) which is not accessible by any current terrestrial experimental facility. At electroweak (EW) scale ($E_{EW}\sim 100$ GeV) the effect of Lorentz violation is suppressed by a factor of $E_{EW}/E_{Pl}\sim 10^{-17}$. However, several experimental techniques have been developed over the years to test the existence of LV. The system of three flavour neutrino oscillation is one of the most commendatory sector to study such effects which can also be understood as interference between different distinguishable states \cite{Cabo:2021exm} . The analysis of LV in the framework of neutrino oscillation has been done previously in several works \cite{Bahcall:2002ia,Antonelli:2018fbv,Torri:2020dec,Super-Kamiokande:2014exs,Kostelecky:2004hg,KumarAgarwalla:2019gdj,Sahoo:2021dit,Arias:2006vgq}. The Lagrangian density corresponding to LV considering the effect of CPT violation is expressed as \cite{Diaz:2015dxa}
\begin{equation}
\mathcal{L}_{LV} \subset a_{\alpha\beta}^{\lambda}(\bar{\nu}_{\alpha}\gamma_{\lambda}P_{L}\nu_{\beta}),
\end{equation}
where $P_L=(1-\gamma_5)/2$ and $a_{\alpha\beta}^{\lambda}$ is the parameter exhibiting the strength of LV. Here $\lambda$ is the spacetime index ($\lambda=0,1,2,3$) and $\alpha,\beta$ are the flavour indices, $\alpha,\beta \in \{e,\mu,\tau\}$. In this analysis, only the isotropic component of LV parameters ($\lambda =0$) are taken into account. The interaction Hamiltonian corresponding to the phenomenon of LV is given by
\begin{equation}
\mathcal{H}_{LV}= \begin{pmatrix}
a_{ee} & a_{e\mu} & a_{e\tau} \\
a_{\mu e} & a_{\mu \mu} & a_{\mu \tau} \\
a_{\tau e} & a_{\tau \mu} & a_{\tau \tau}
\end{pmatrix}    
\end{equation}
Here $a_{\alpha\beta}^{\lambda=0}$ is simply denoted by $a_{\alpha\beta}$. As $\mathcal{H}_{LV}$ is Hermitian, $a_{\alpha\beta}^*=a_{\beta\alpha}$. Due to the consideration of CPT-violating scenario, $a^{\lambda}_{\alpha\beta} \xrightarrow[\text{}]{\text{CPT}} -a^{\lambda}_{\alpha\beta}$. The off-diagonal elements are generally complex, $a_{\alpha\beta}=|a_{\alpha\beta}|e^{i\phi_{\alpha\beta}}$.
The constraints on these parameters are obtained from various experiments such as IceCube \cite{PhysRevD.82.112003}, MINOS \cite{MINOS:2010kat}, Double Chooz \cite{PhysRevD.86.112009}.

\subsection{Dark non standard interaction}\label{NSI}
In neutrino sector another way to introduce the BSM physics is via the inclusion of the non standard interaction (NSI) which appears in several mass models. Considering SM as a lower energy approximation of a full theory existing at a much higher energy scale ($\Lambda$), the effective Lagrangian is possible to be represented in terms of a number of higher dimensional Lorentz structures. Each of the operator is non-renormalizable, in general, and is restricted by the NP energy scale, $\Lambda$ at which the phenomenon of LV might also occur. 

NSI is exchanged either by a vector or a scalar mediator. The vector mediator is either charged or neutral in nature. The charged current (CC) vector NSI is tightly constrained, while the constraints on the neutral current (NC) vector NSI is much weaker. The importance of NC NSI has been discussed in several literatures \cite{Farzan:2017xzy,Antusch:2008tz,Proceedings:2019qno}. The NC NSI contributes a subleading effect to the neutrino oscillation and the Hamiltonian corresponding to it is given by
\begin{equation}
\mathcal{H}_{NC-NSI}^{vec} = A\begin{pmatrix}
\epsilon_{ee} & \epsilon_{e \mu} & \epsilon_{e \tau} \\
\epsilon_{\mu e} & \epsilon_{\mu\mu} & \epsilon_{\mu \tau} \\
\epsilon_{\tau e} & \epsilon_{\tau \mu} & \epsilon_{\tau \tau}
\end{pmatrix}
\end{equation}
The parameters, $\epsilon_{\alpha\beta}$ determine the strength of NSI which corresponds to the dimension-$6$ four fermion lepton number conserving operator. The off-diagonal elements ($\alpha \neq \beta$) are complex, in general. $A=\pm G_F n_e$, with $G_F$ and $n_e$ representing Fermi constant and the number density of electron in the medium respectively. The $'+'$ and $'-'$ signs correspond to neutrino and anti-neutrino respectively. In a recent analysis, the degeneracy existing between the scenario of LV and vector NSI is discussed along with the role of atmospheric neutrino to lift up this degeneracy \cite{sahoo2022core}.

While the vector NSI affects the matter potential during neutrino propagation through a medium, the NSI mediated by a scalar particle has completely different consequences. Instead of modifying the SM matter potential, the scalar NSI directly influences the neutrino mass term. The effect of scalar mediator is added up to the mass term as a perturbation. 
According to the modern cosmological model, the universe is filled with dark matter (DM) particles ($\sim 27 \%$). From several astrophysical constraints the local dark matter density is estimated to be $\rho_{\chi}\sim 0.47$ GeV/cm$^3$. Since the number density of the DM particles is inversely proportional to its mass, there can be a large abundance of DM present in the universe if their mass is sufficiently small. These light DM particles ($m_{\chi} \leq 100$ eV) can be either scalar or vector, and they may also interact with neutrinos. Such interaction generates another kind of NSI, named as dark NSI. If the DM mediator is a scalar particle ($\phi$) of mass $m_{\phi}$, the neutrino-DM interaction Lagrangian is given by \cite{Ge:2019tdi,article}
\begin{equation}
-\mathcal{L}_{NSI}^{dark} \subset \frac{1}{2} M_{\alpha\beta}\bar{\nu}_{\alpha}\nu_{\beta}+y_{\alpha\beta}\bar{\nu}_{\alpha}\phi \nu_{\beta} +\frac{1}{2}m_{\phi}^2 \phi^2     
\end{equation}
Here $y_{\alpha\beta}$ is the Yukawa coupling corresponding to neutrino-DM interaction and $M_{\alpha \beta}$ is the neutrino mass matrix. The corresponding interaction Hamiltonian is given by \cite{Ge:2019tdi,article}
\begin{equation}
\mathcal{H}_{NSI}^{dark}= \frac{1}{2E_{\nu}} (MM^{\dagger}+\delta M^2)    
\end{equation}
Here the correction to the mass term is represented as, $\delta M^2_{\alpha\beta}=\frac{2\rho_{\phi}}{m_{\phi}^2}y_{\alpha j}y^*_{j\beta}$. Such coupling of neutrino with DM is shown to violate the CPT symmetry \cite{Ge:2019tdi}. The correction to the mass term due to the presence of dark NSI is parametrized as follows \cite{Ge:2018uhz, Medhi:2021wxj}
\begin{equation}
\delta M \equiv \sqrt{\Delta m_{31}^2} \begin{pmatrix}
\eta_{ee} & \eta_{e\mu} & \eta_{e\tau} \\
\eta_{\mu e} & \eta_{\mu \mu} & \eta_{\mu \tau} \\
\eta_{\tau e} & \eta_{\tau \mu} & \eta_{\tau\tau}.
\end{pmatrix}
\end{equation}
Here $\eta_{\alpha \beta}$ determines the strength of dark NSI. From the recent data obtained from Borexino $2017$, the constraint on only one diagonal NSI parameter is estimated 
\cite{Ge:2018uhz}.

In the next section, the analytical description to obtain neutrino oscillation probability is discussed in the scenario of LV and dark NSI with the consideration of unitary and non-unitary neutrino mixing matrix.

\subsection{Neutrino oscillation in matter}\label{matter}
The phenomenon of neutrino oscillation is based on the fact that neutrino flavour eigen states are represented as a superposition of the mass eigen states. Considering an initial neutrino beam $\nu_{\alpha}$ ($\alpha=e,\mu,\tau$), the relation between the flavour and the mass eigen states is represented as
\begin{equation}
\ket{\nu_{\alpha}(t=0)}=\sum_{i=1}^3 U^{3\times 3*}_{\alpha i}\ket{\nu_i(t=0)} 
\end{equation}
The matrix $U^{3\times 3}$ might be unitary ($U_{PMNS}$) or  non-unitary ($T^{3\times 3}$), represented by eqn. (2) and (3) respectively. 
Now, the time evolution of the mass eigen state is expressed as, $\ket{\nu_i(t)}=e^{-iH_m t} \ket{\nu_i(t=0)} = e^{-iE_i t}$, where $H_m$ is the interaction Hamiltonian in mass basis and $E_i$ is the corresponding energy eigen value. In mass basis the effective Hamiltonian in presence of Lorentz violation (LV) is represented as
\begin{eqnarray}
H_m = \frac{1}{2E_{\nu}}\begin{pmatrix} 
0 & 0 & 0 \\
0 & \Delta m_{21}^2 & 0 \\
0 & 0 & \Delta m_{31}^2
\end{pmatrix}  + U^{3\times 3} A\begin{pmatrix}
1 & 0 & 0 \\
0 & 0 & 0 \\
0 & 0 & 0
\end{pmatrix} U^{3\times 3 \dagger}\nonumber \\
+ U^{3\times 3}\begin{pmatrix}
a_{ee} & a_{e\mu} & a_{e\tau} \\
a_{\mu e} & a_{\mu \mu} & a_{\mu \tau} \\
a_{\tau e} & a_{\tau \mu} & a_{\tau \tau} 
\end{pmatrix}U^{3\times 3\dagger}
\end{eqnarray}
while considering the effect of dark NSI, the effective Hamiltonian is modified as
\begin{eqnarray}
H_m =  \frac{1}{2E_{\nu}}\begin{pmatrix} 
0 & 0 & 0 \\
0 & \Delta m_{21}^2 & 0 \\
0 & 0 & \Delta m_{31}^2
\end{pmatrix}  + U^{3\times 3} A\begin{pmatrix}
1 & 0 & 0 \\
0 & 0 & 0 \\
0 & 0 & 0
\end{pmatrix} U^{3\times 3 \dagger}\nonumber \\+ \frac{1}{2E_{\nu}} (MM^{\dagger}+\delta M^2)
\end{eqnarray}

\begin{figure*}[ht!]
    \centering
    \includegraphics[scale=0.8]{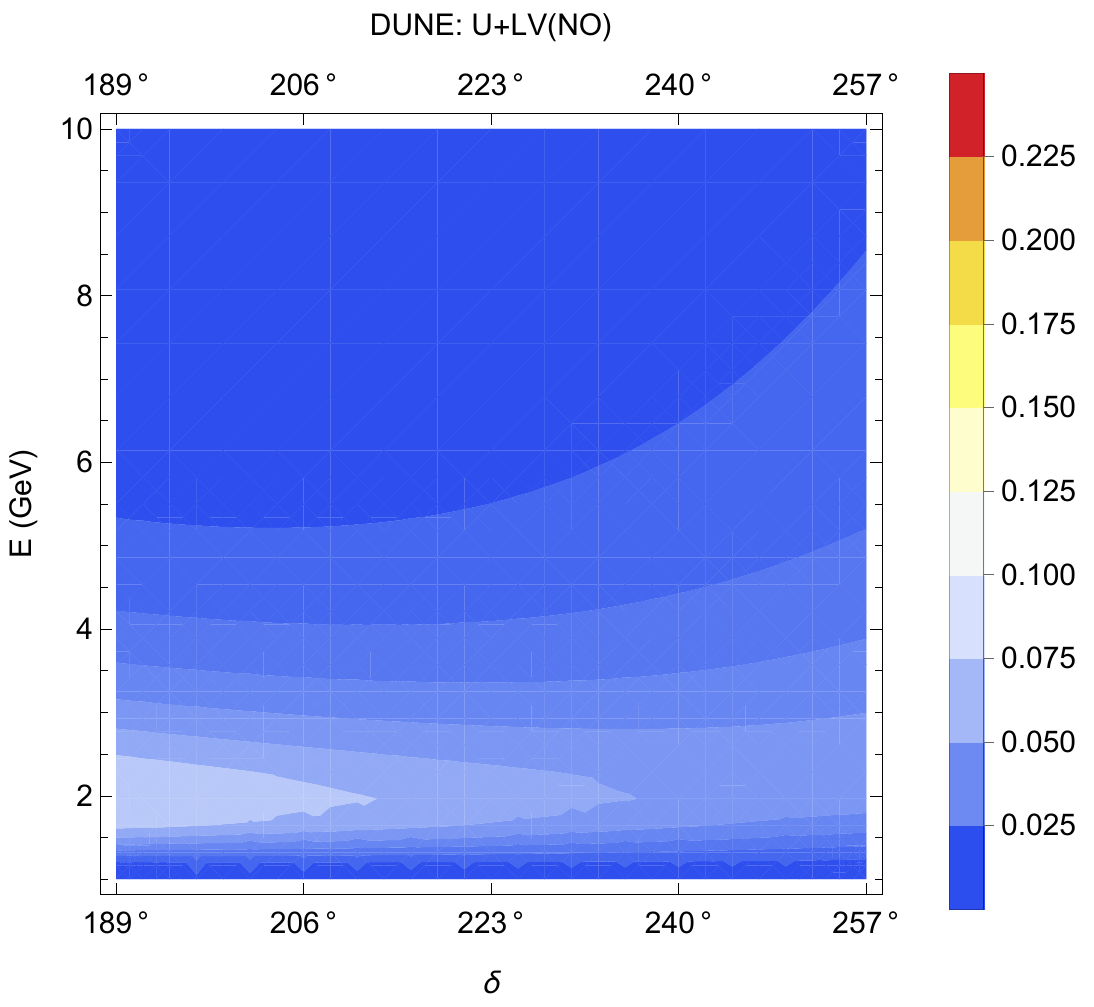}\hspace*{1mm} \includegraphics[scale=0.8]{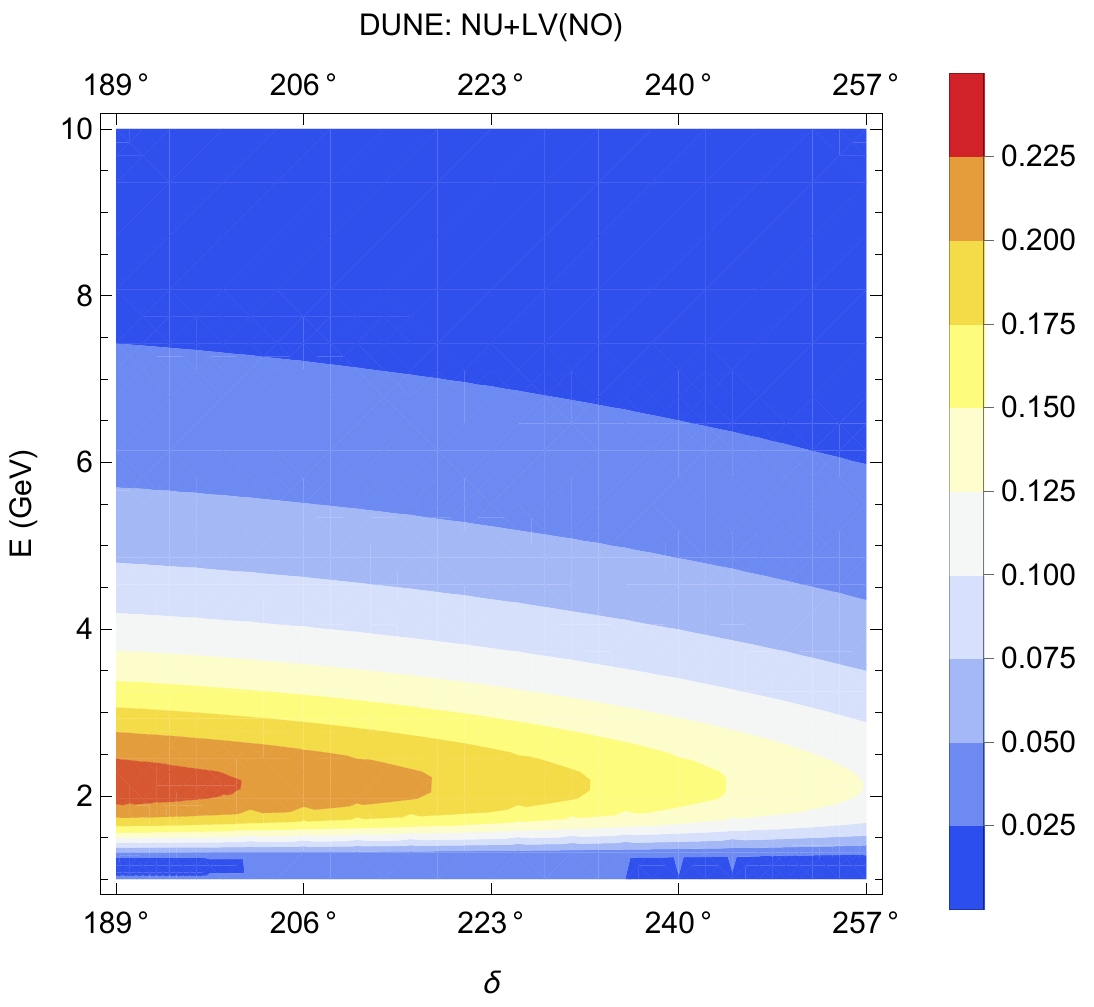} 
    \caption{Variation of oscillation probability $P_{\mu e}$ in ($E_{\nu},\delta$) plane in the scenario of Lorentz violation (LV), in the context of DUNE experiment ($E_{\nu}=1-10$ GeV, $L=1300$ km) for NO, considering unitary (left panel) and non-unitary (right panel) neutrino mixing.}\label{f1}
\end{figure*}

\begin{figure*}[ht!]
    \centering
     \includegraphics[scale=0.8]{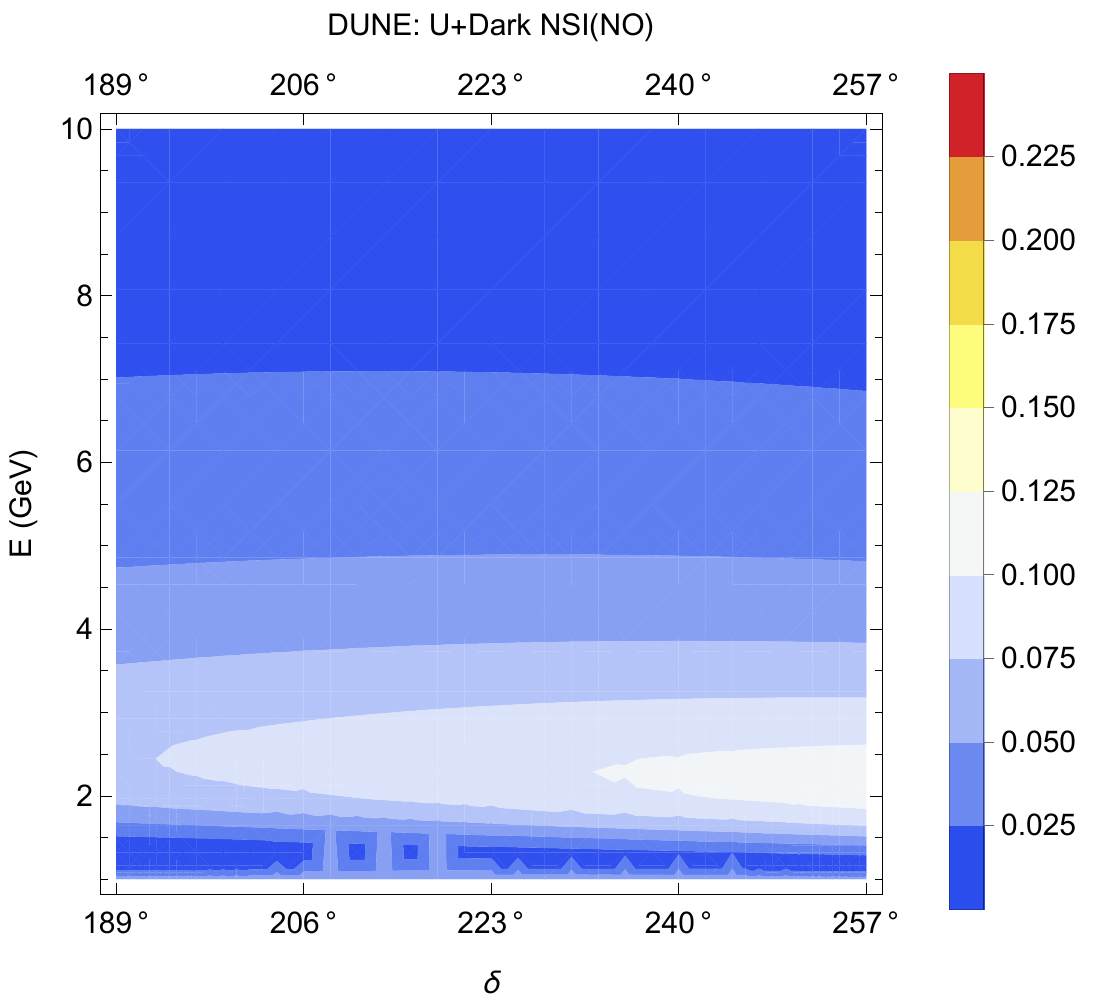}\hspace*{1mm} \includegraphics[scale=0.8]{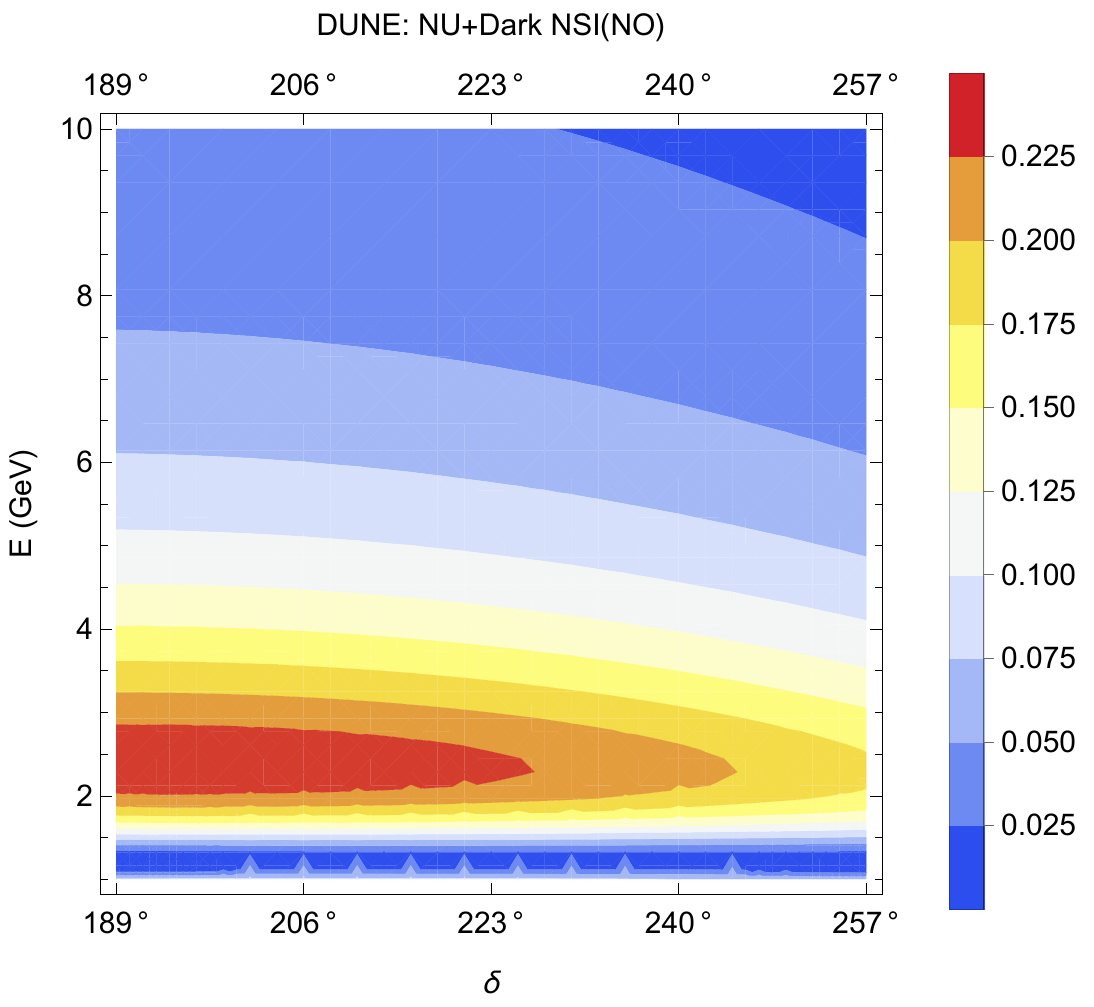} 
    \caption{Same as fig. \ref{f1} in the scenario of dark NSI.}
    \label{f2}
\end{figure*}

\begin{figure*}[ht!]
    \centering
    \includegraphics[scale=0.8]{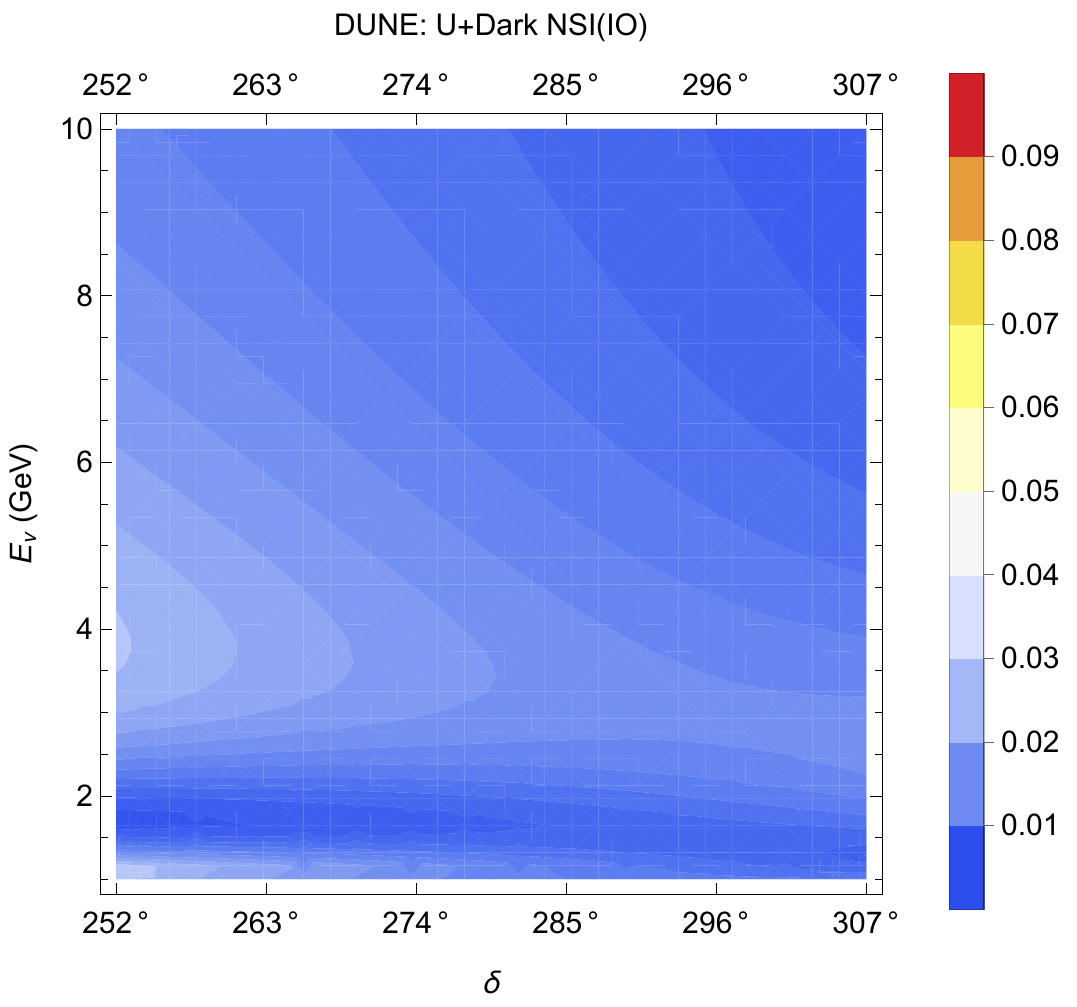}\hspace*{1mm} \includegraphics[scale=0.8]{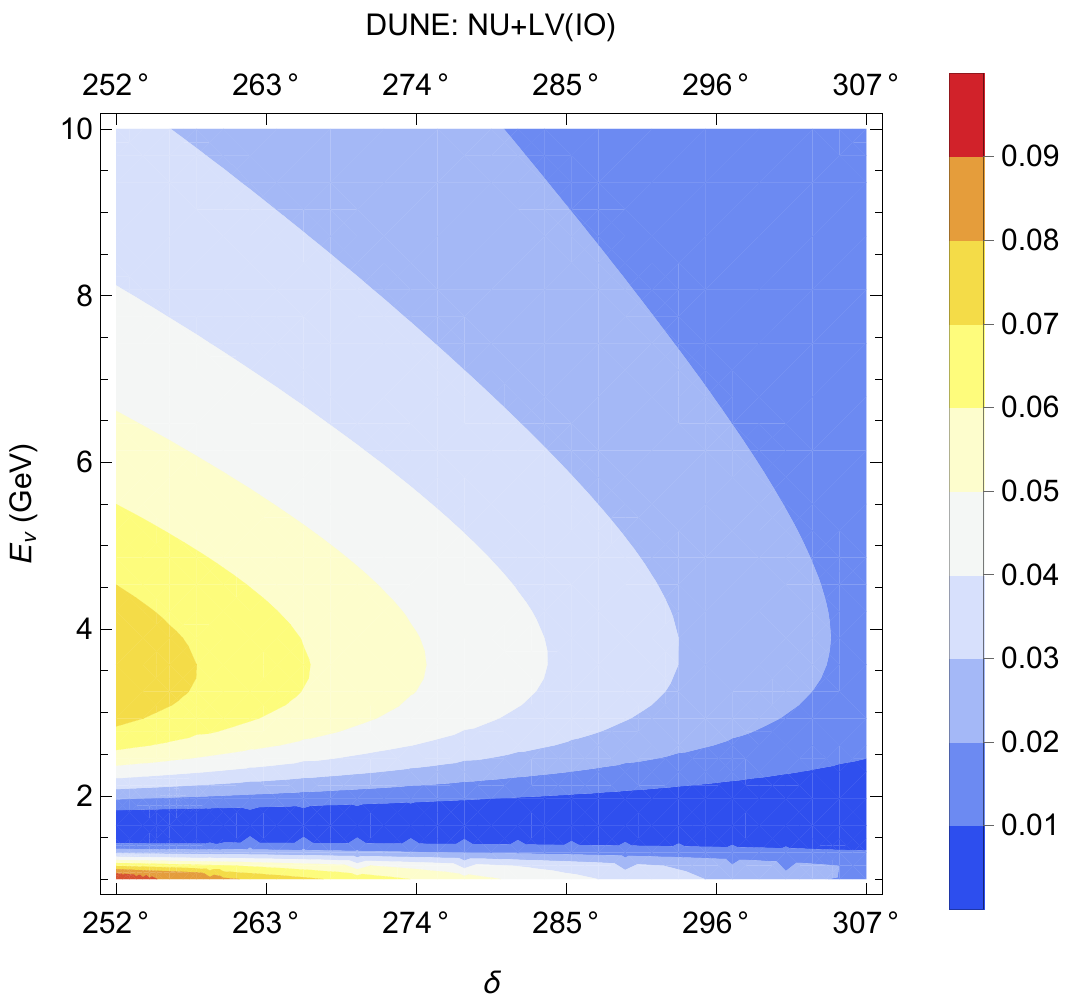}
    \caption{Variation of oscillation probability $P_{\mu e}$ in ($E_{\nu},\delta$) plane in the scenario of Lorentz violation (LV), in the context of DUNE experiment ($E_{\nu}=1-10$ GeV, $L=1300$ km) for IO, considering unitary (left panel) and non-unitary (right panel) neutrino mixing.}\label{f5}
    \end{figure*}
    
\begin{figure*}    
    \includegraphics[scale=0.8]{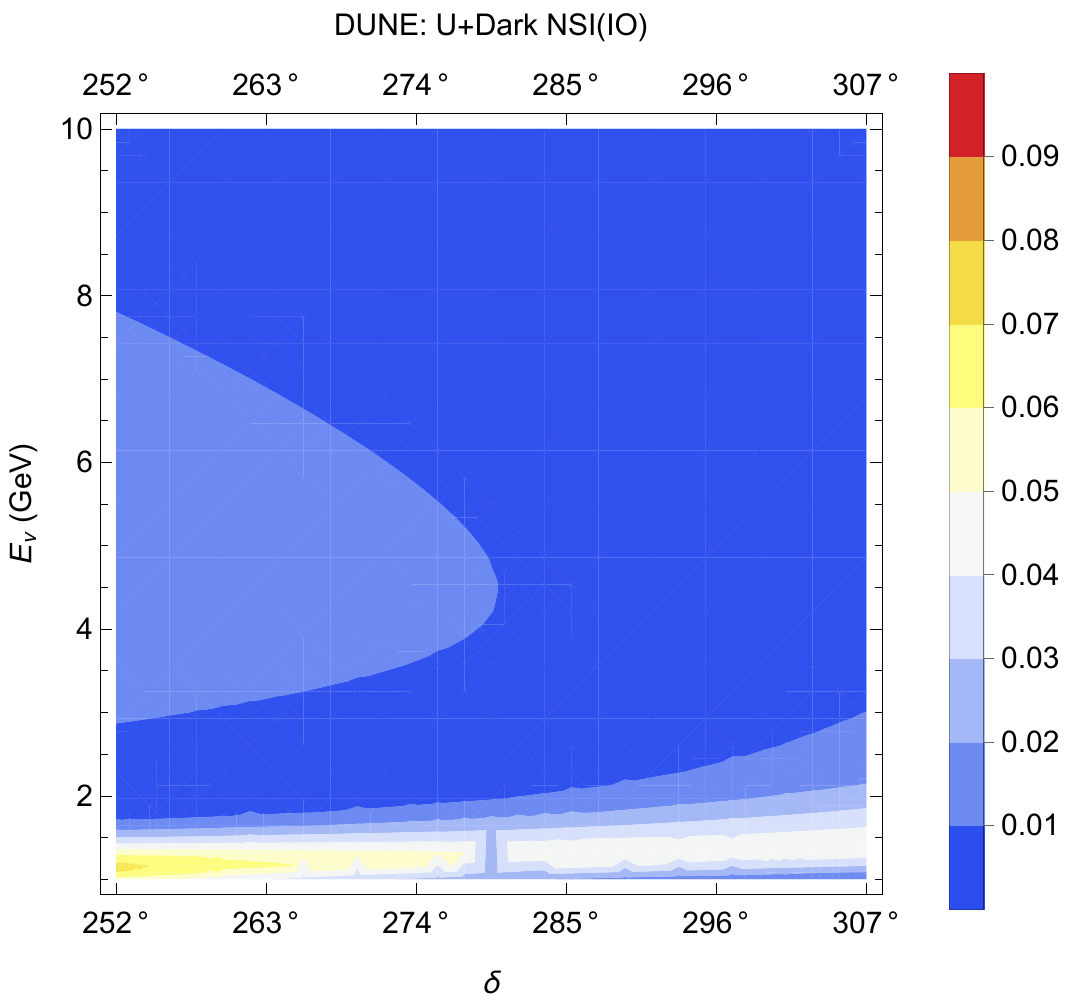}\hspace*{1mm} \includegraphics[scale=0.8]{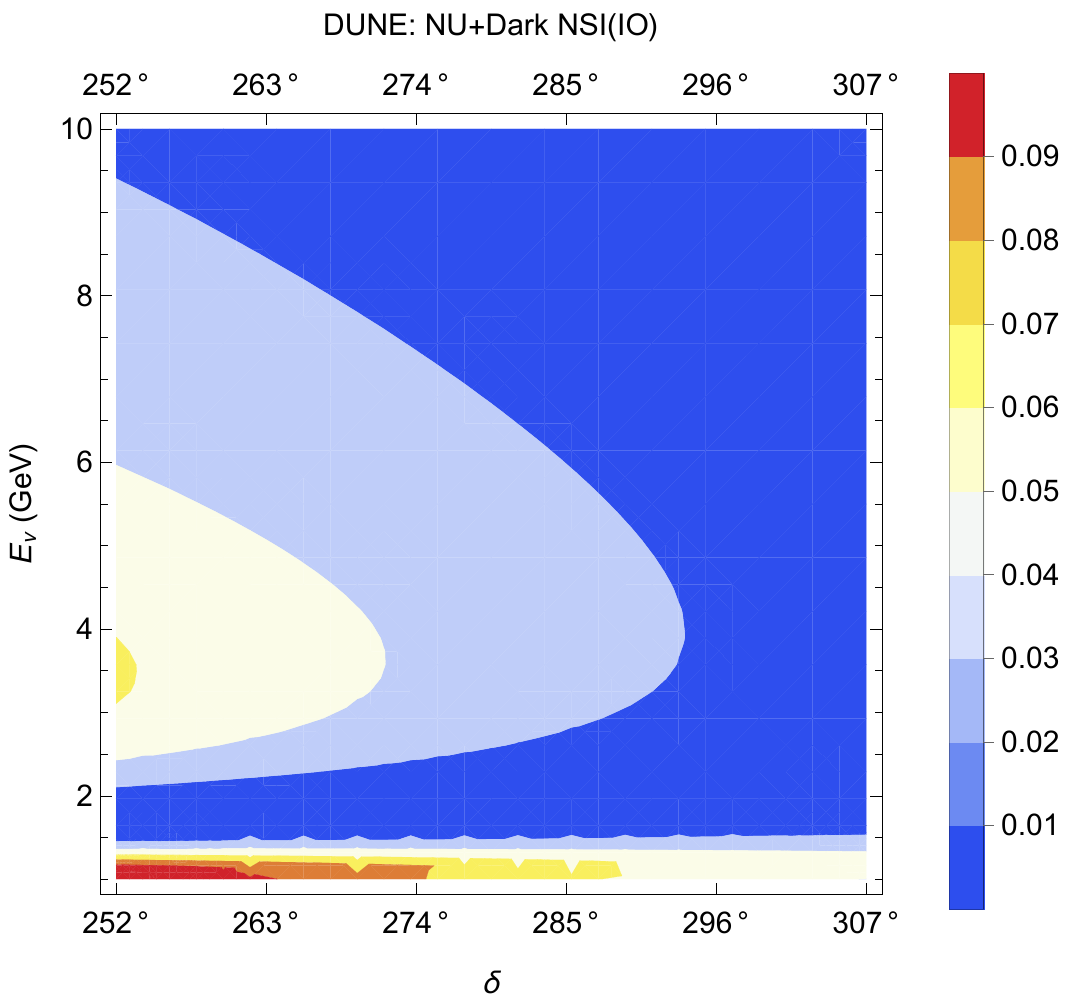}
      \caption{Same as fig. \ref{f5} in case of dark NSI.}
    \label{f6}
\end{figure*}

The unperturbed part of the mass matrix $M$ is given by, $M\equiv U^{3\times 3}.Diag(m_1,m_2,m_3). U^{3\times 3\dagger}$. The time evolution of the flavour state is expressed as, $\ket{\nu_{\alpha}(t)}=e^{-iH_f t}\ket{\nu_{\alpha}(t=0)}$, $H_f$ is the Hamiltonian in the flavour basis, $H_f=U^{3\times 3}H_m U^{3\times 3\dagger}$. To calculate the probability of neutrino oscillation, the framework of ref. \cite{Ohlsson:1999xb} is taken into consideration, according to which the flavour evolution operator at ultra relativistic limit is expressed as 
\begin{eqnarray}
U_f(L)=e^{-iH_f L} = \phi \sum_{i=1}^3 e^{-iL\lambda_a} \frac{1}{3\lambda_a^2+C_1} \times \nonumber \\
\left[(\lambda_a^2+C_1)I_3 +\lambda_a \Tilde{T}+\Tilde{T}^2\right]
\end{eqnarray}
Here $T=H_m-tr(H_m)I_3/3$, $I_3$ being $3\times 3$ identity matrix. The matrix, $T$ is $3\times 3$ traceless  Hermitian matrix with eigenvalues $\lambda_a$. $\phi=exp[-iLtr(H_m)I_3/3]$, $\Tilde{T}=U^{3\times 3}TU^{3\times 3 \dagger}$ and $C_1=Det(T)tr(T^{-1})$.
\begin{center}
\begin{table}
\begin{tabular}{|c |c| c|} 
 \hline
Parameters & Best fit$\pm~1\sigma$ (NO) & Best fit$\pm~1\sigma$ (IO) \\ [0.5ex] 
 \hline\hline
$\theta_{12}^o$ & $33.82^{+0.78}_{-0.76}$ & $33.82^{+0.78}_{-0.75}$\\
 \hline
$\theta_{23}^o$ & $49.7^{+0.9}_{-1.1}$ &  $49.7^{+0.9}_{-1.0}$\\
\hline
$\theta_{13}^o$ & $8.61^{+0.12}_{-0.13}$ &  $8.65^{+0.12}_{-0.13}$\\
\hline
$\delta$ & $217^{+40}_{-28}$ & $280^{+25}_{-28}$ \\
\hline
$\Delta m_{21}^2$ (eV$^2$) & $7.39^{+0.21}_{-0.20}\times 10^{-5}$ & $7.39^{+0.21}_{-0.20}\times 10^{-5}$\\
\hline
$\Delta m_{3l}^2$ (eV$^2$) & $+2.525^{+0.033}_{-0.031}\times 10^{-3}$ & $-2.512^{+0.034}_{-0.031}\times 10^{-3}$\\
\hline
\end{tabular}
\caption{Values of neutrino oscillation parameters \cite{Esteban:2018azc} with $1\sigma$ interval}\label{t1}
\end{table}
\end{center}

\begin{center}
\begin{table}
\begin{tabular}{|c | c|} 
 \hline
 Parameters & Best fit with $3\sigma$ interval \\
 \hline
$\alpha_{00}$ & $0.93$ \\
\hline
$\alpha_{11}$ & $0.95$ \\
\hline
$\alpha_{22}$ & $0.95$ \\
\hline
$|\alpha_{10}|$ & $3.6\times 10^{-2}$\\
\hline
$|\alpha_{20}|$ & $0.13$ \\
\hline
$|\alpha_{21}|$ & $0.021$ \\
\hline
\end{tabular}
\caption{Non-unitarity parameters \cite{Miranda:2019ynh} with $3\sigma$ interval. The phase factors are considered to be zero.}
\label{tt2}
\end{table}
\end{center}

\section{Results and Discussion}\label{result}
\begin{figure*}[ht!]
    \centering
     \includegraphics[scale=0.8]{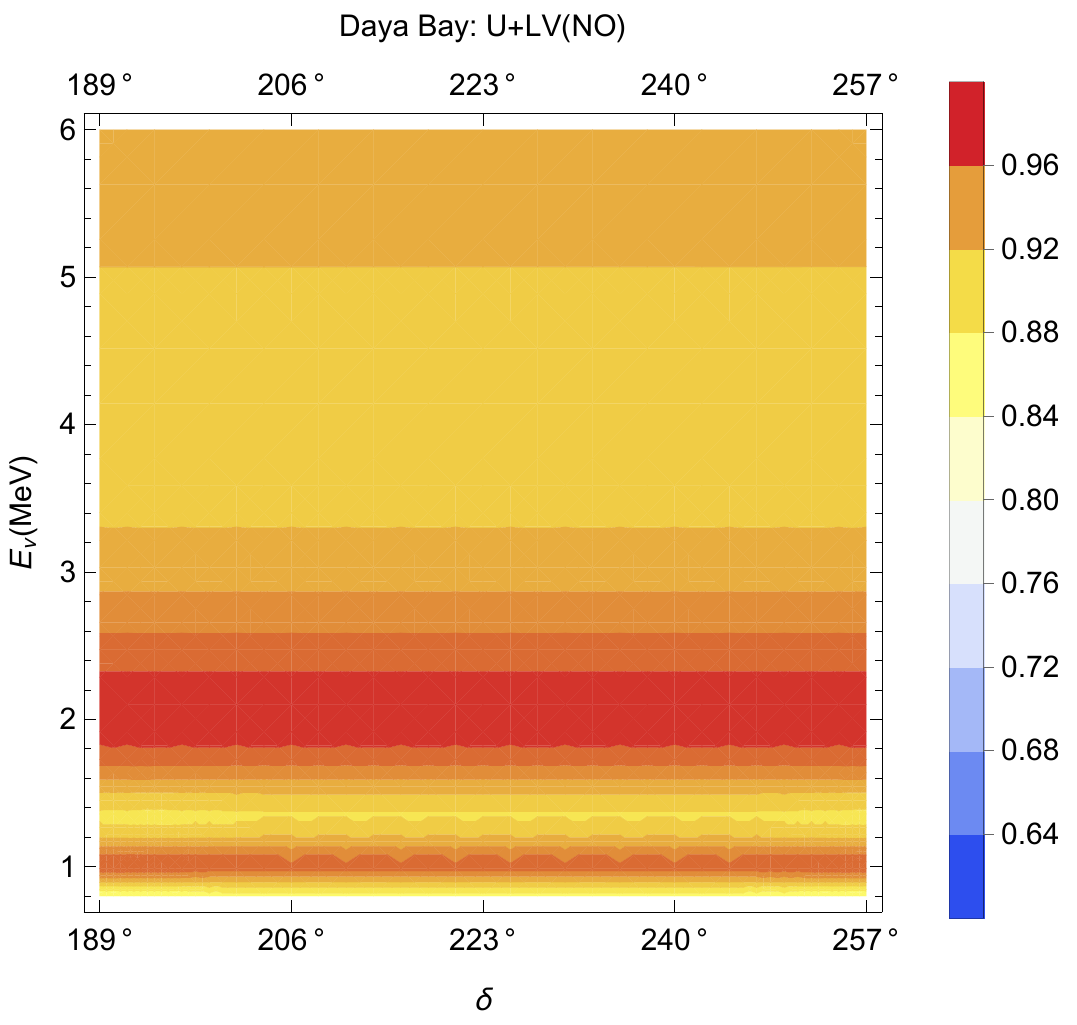}\hspace*{1mm} \includegraphics[scale=0.8]{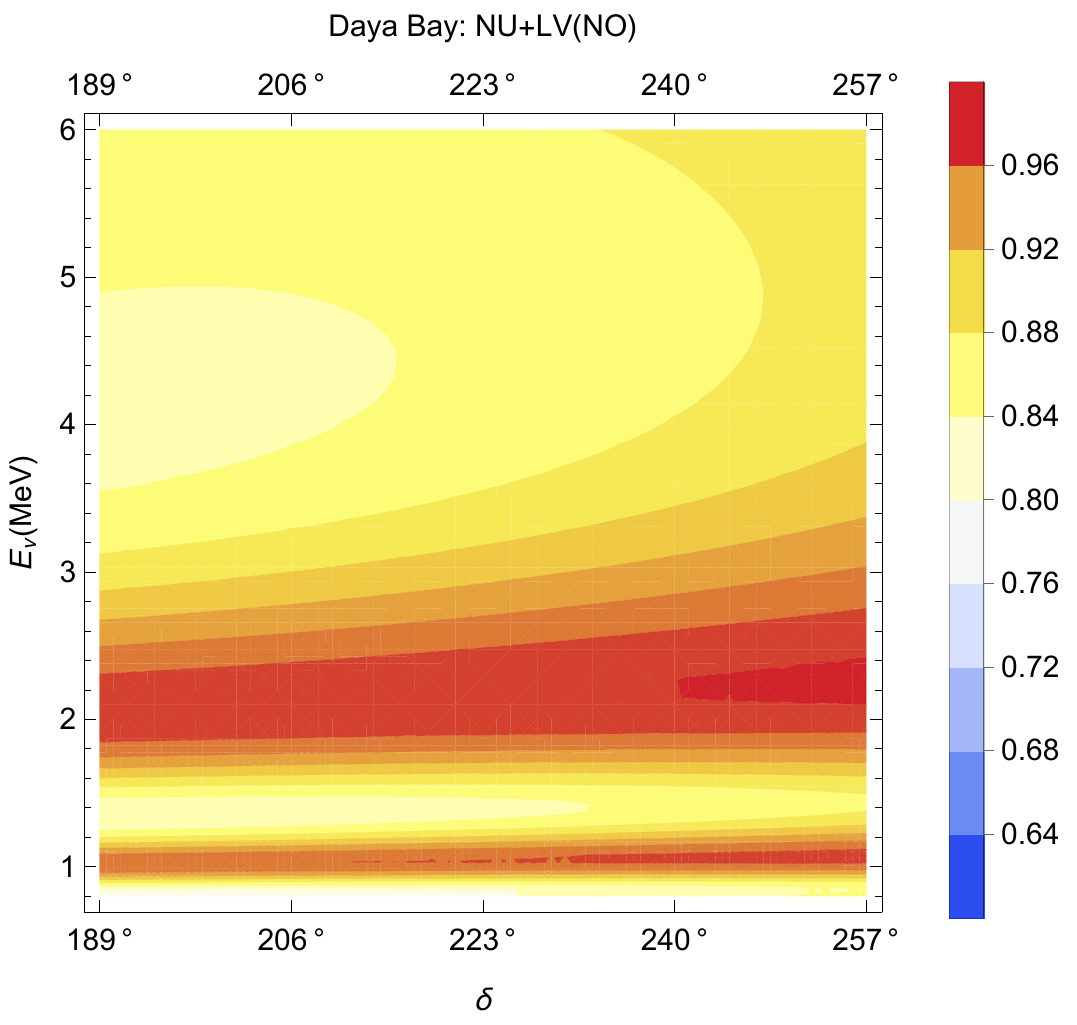}
\caption{Variation of $\bar{\nu}_{ee}$ survival probability $\bar{P}_{ee}$ in ($E_{\nu},\delta$) plane in the scenario of Lorentz violation (LV), in the context of Daya Bay experiment ($E_{\nu}=0.8-6$ MeV, $L=2$ km) for NO, considering unitary (left panel) and non-unitary (right panel) neutrino mixing.}\label{f} 
\end{figure*}

\begin{figure*}
\centering
   \includegraphics[scale=0.8]{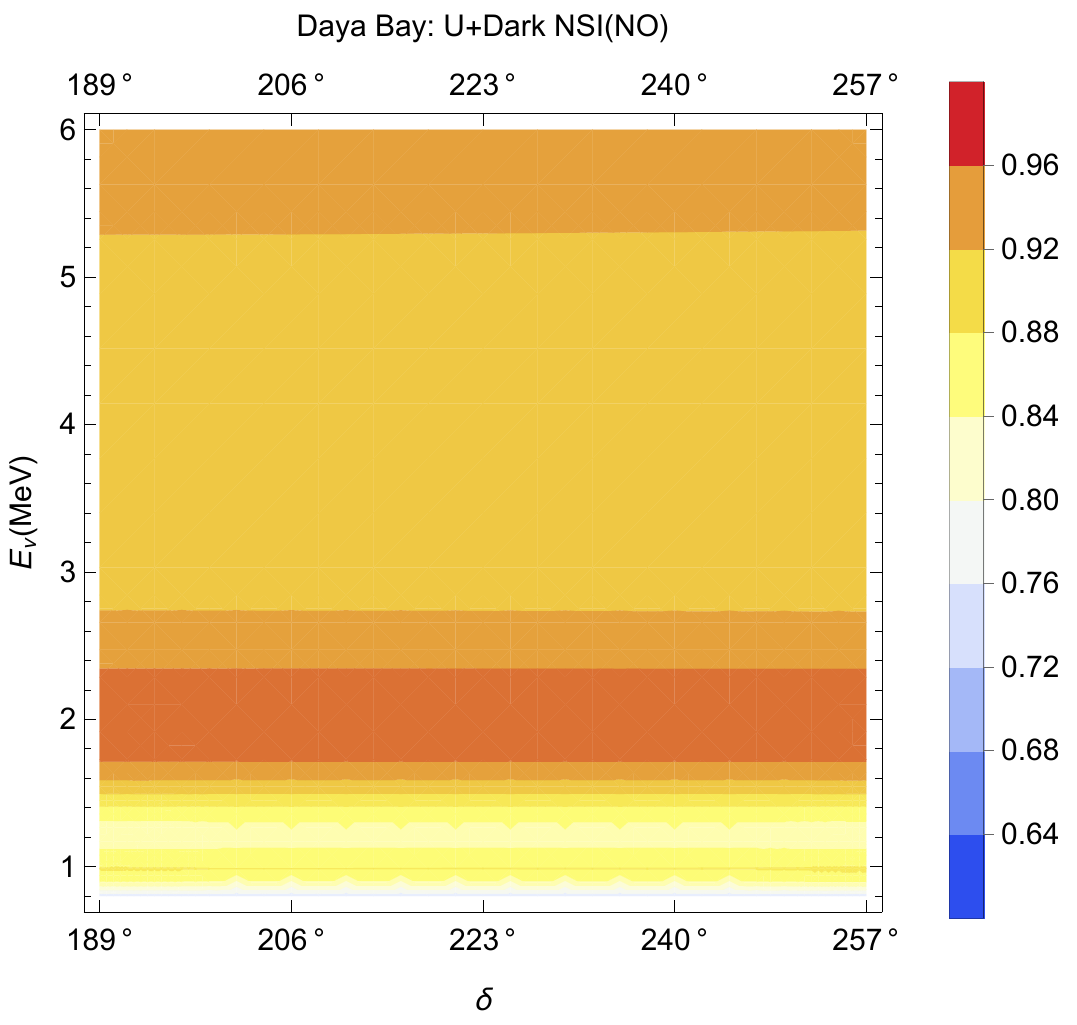}\hspace*{1mm} \includegraphics[scale=0.8]{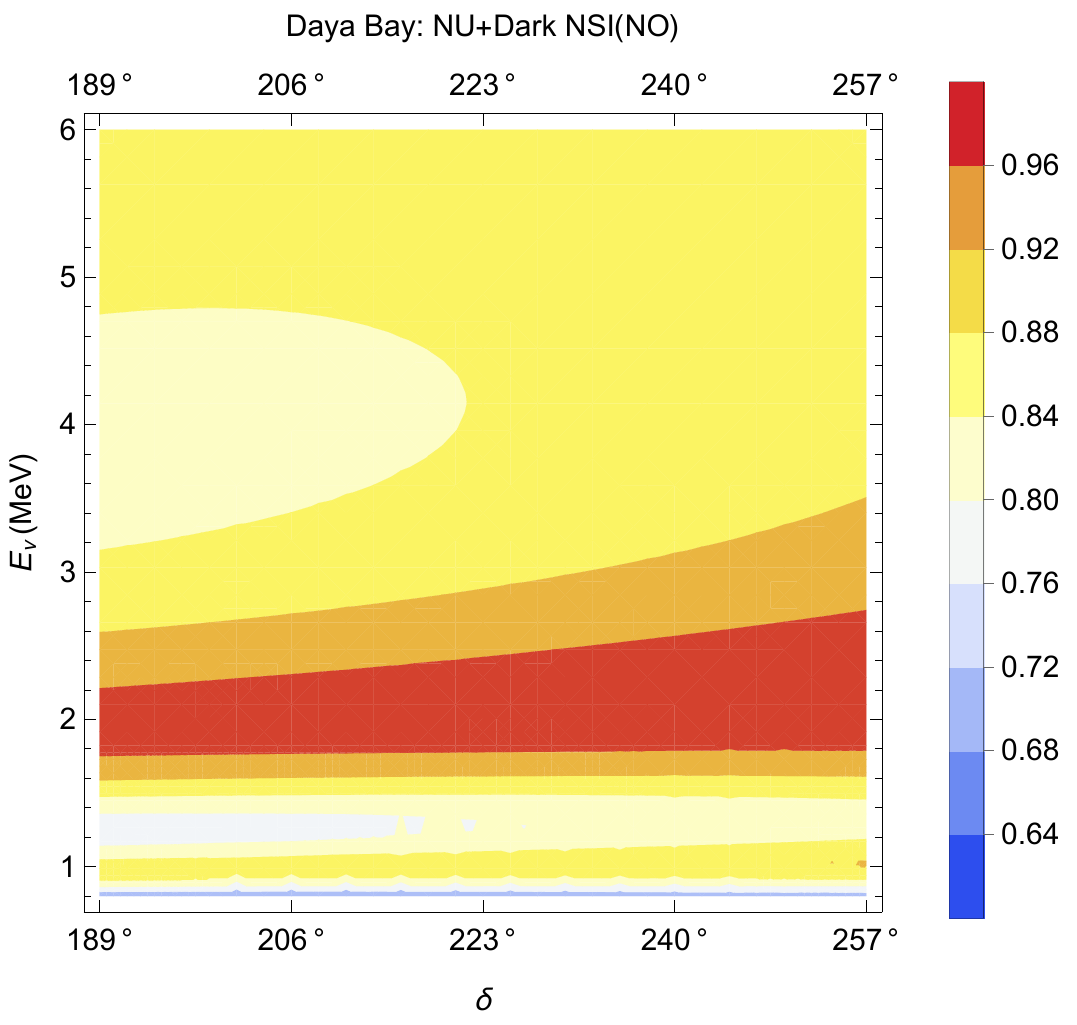}
    \caption{Same as fig. \ref{f} in case of dark NSI}\label{ff}
\end{figure*}

In this section the effects of unitary and non unitary mixing matrix in the neutrino  oscillation and survival probability are analysed and compared for both normal (NO) and inverted mass ordering (IO), considering two of the new physics scenarios, LV and dark NSI, in the context of DUNE and Daya Bay experimental set ups respectively. 

($1$) DUNE is an LBL accelerator experiment having the baseline length $L=1300$ km, which operates within the energy range $E_{\nu}=1-10$ GeV. The oscillation channel mainly observed in DUNE is $\nu_{\mu}\rightarrow \nu_e$ and the maximum flux is observed around the energy, $E_{\nu}\sim 2.5-3$ GeV \cite{Brailsford:2018dzn}.

($2$) Daya Bay is an SBL reactor experiment with baseline length of $L=2$ km operating within the energy range $E_{\nu}=0.8-6$ MeV which observes the $\bar{\nu}_e$ beam and measures its survival probability \cite{DayaBay:2012fng}. It should be mentioned that at $E_{\nu}\approx 4-5$ MeV an excess of flux is observed which is larger than the expected value predicted from the theory \cite{an2016measurement}. Such discrepancy is also observed at $E_{\nu}\approx 1$ MeV in a much lesser extent \cite{an2016measurement}. Further, it should be be mentioned that at $E_{\nu}\sim 1$ MeV the energy resolution of the detector is $\sim 8.5\%$ \cite{an2021antineutrino}.

 The Earth matter density considered in this work is $\rho=2.8$ gm/c.c. which generates the matter potential $A=1.01\times 10^{-13}$ eV. The mass eigen values of the three neutrinos are considered as, $m_1=10^{-5}$ eV, $m_2=\sqrt{\Delta m_{21}^2+m_1^2},~m_3=\sqrt{\Delta m_{31}^2+m_2^2}$. In this analysis only one diagonal NSI parameter is considered to be non zero, given by  $\eta_{ee}\in[-0.3,0.3]$ \cite{Medhi:2021wxj}. The LV parameter is chosen as, $a_{e\mu}=10^{-24}$ GeV with the corresponding complex phase, $\phi_{e\mu}=0$, while the rest of the parameters are taken to be zero.  The standard neutrino oscillation parameters in case of normal (NO) and inverted mass ordering (IO) are illustrated in table \ref{t1} \cite{Esteban:2018azc}. The bounds on NU parameters ($\alpha_{ij}$) are presented in table \ref{tt2}.
\begin{center}
\begin{table}[h!]
\begin{tabular}{|c|c|c|c||c|c|} 
\hline
\multicolumn{6}{|c|}{DUNE} \\
\hline
 \multirow{2}{*}{\vtop{\hbox{\strut Mass}\hbox{\strut ordering}}} & ($E_{\nu}$,$\delta$) & \multicolumn{2}{c||}{Unitary (U)} & \multicolumn{2}{c|}{Non-unitary (NU)}\\
  \cline{3-6}
 & & LV & NSI &  LV & NSI  \\
  \hline
  \multirow{2}{*}{NO} & $E_{\nu}$ (GeV) &$1.5-2.5$ & $2-2.5$ & $2-2.5$ & $2-2.75$  \\
  \cline{2-6}
 & $\delta$ ($^o$) & $189-215$ & $189-200$ & $230-257$ & $189-230$ \\
  \hline
 \multirow{2}{*}{IO} & $E_{\nu}$ (GeV) & $1$, $3.5-4$ & $1$ & $1$ & $1$  \\
  \cline{2-6}
 & $\delta$ ($^o$) & $252$ & $252$ & $252$ & $252$ \\
  \hline
\end{tabular}
\caption{The specific ranges of the neutrino energy $E_{\nu}$ and CP violating phase $\delta$ corresponding to $P_{\mu e,max}\gtrsim 0.225$ in case of NO and $P_{\mu e,max}\gtrsim 0.09$ in case of IO, for the two NP scenarios of LV and dark NSI, under the assumption of unitary and non-unitary neutrino mixing, in context of LBL DUNE experimental set up ($E_{\nu}=1-10$ GeV, $L=1300$ km).}\label{tt3}
\end{table}
\end{center}

In fig. \ref{f1} and \ref{f2}, the variation of the transition probability in the $\nu_{\mu}\rightarrow \nu_e$ channel is depicted in the plane of ($E_{\nu},\delta$) in the scenarios of LV and dark NSI respectively, for NO in the context of DUNE experimental set up. It can be seen from the figure that $P_{\mu e}$ attains large value ($\gtrsim 0.225$), with the consideration of NU mixing only. In the LV scenario, $P_{\mu e}\gtrsim 0.225$ spans the region confined within $189^o\lesssim \delta \lesssim 200^o$, corresponding to the energy range $E_{\nu}\approx 2-2.5$ GeV, while in case of the dark NSI, this is further extended to a wider region of $(E_{\nu},\delta)$ plane $i.e.$ $189^o\lesssim \delta \lesssim 230^o$ for the energy interval $E_{\nu}\approx 2-2.8$ GeV. However, for the unitary neutrino mixing $P_{\mu e}$ does not attain such a large value and exists below the limit $\sim 0.125$ throughout the entire plane.

\begin{figure*}[ht!]
    \centering
    \includegraphics[scale=0.8]{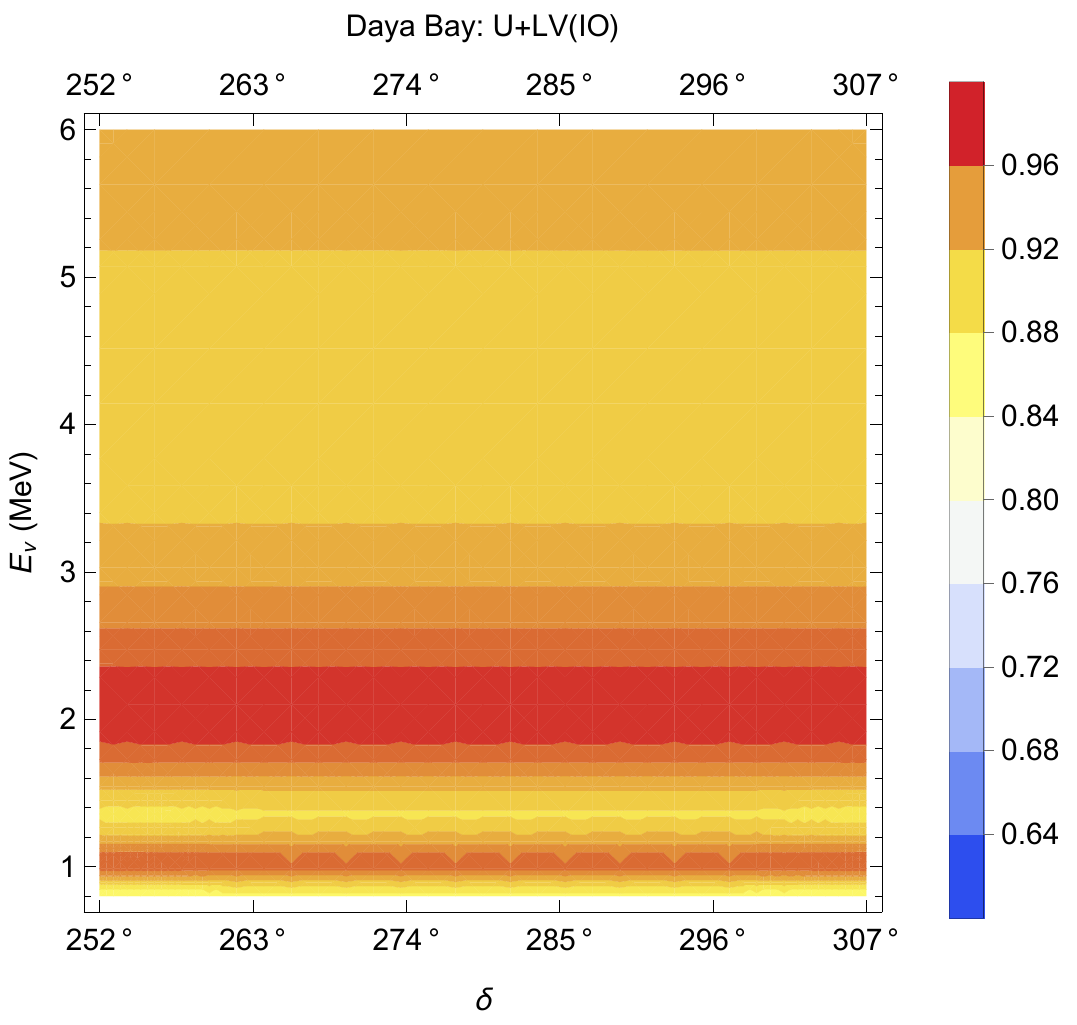}\hspace*{1mm} \includegraphics[scale=0.8]{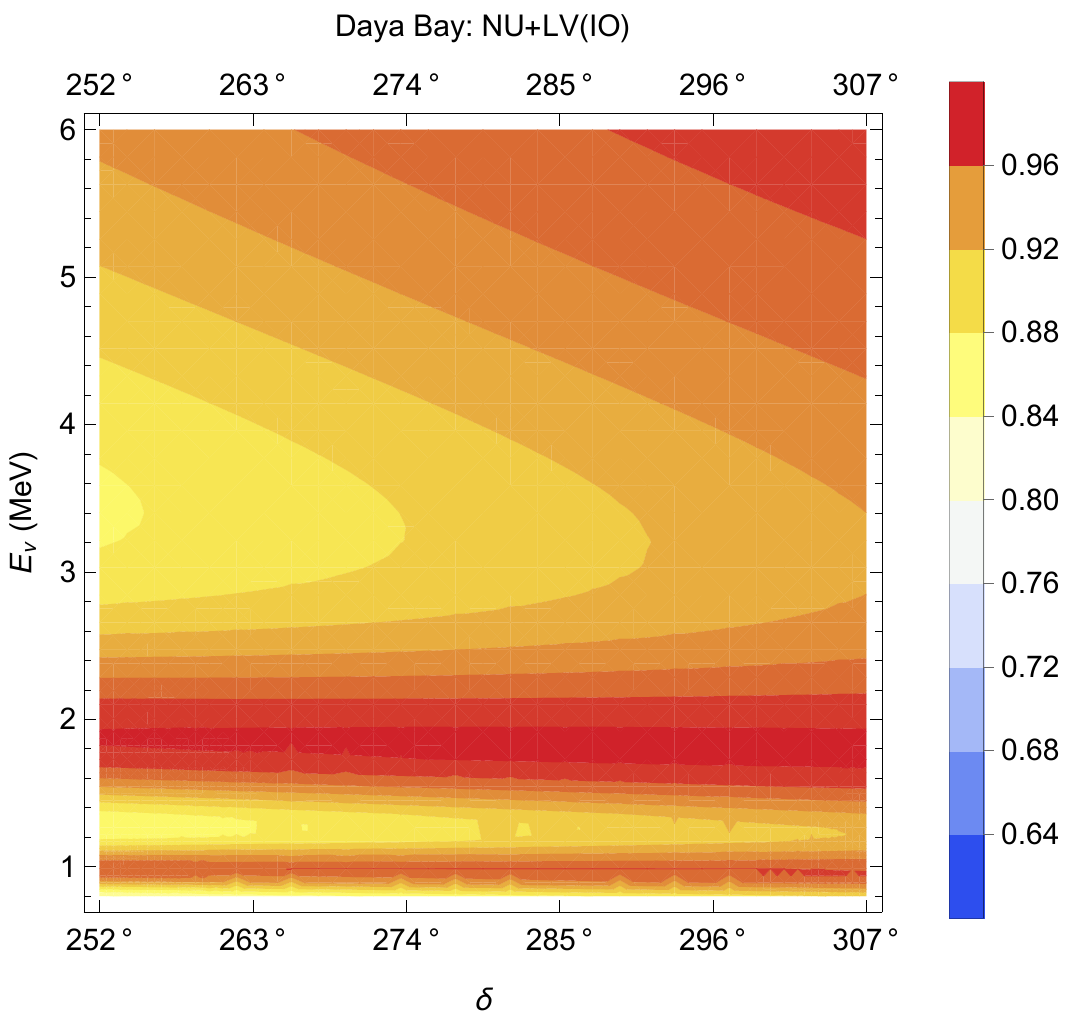}
\caption{Variation of $\bar{\nu}_{ee}$ survival probability $\bar{P}_{ee}$ in ($E_{\nu},\delta$) plane in the scenario of Lorentz violation (LV), in the context of Daya Bay experiment ($E_{\nu}=0.8-6$ MeV, $L=2$ km) for IO, considering unitary (left panel) and non-unitary (right panel) neutrino mixing.}\label{f7}
\end{figure*}

\begin{figure*}
     \includegraphics[scale=0.8]{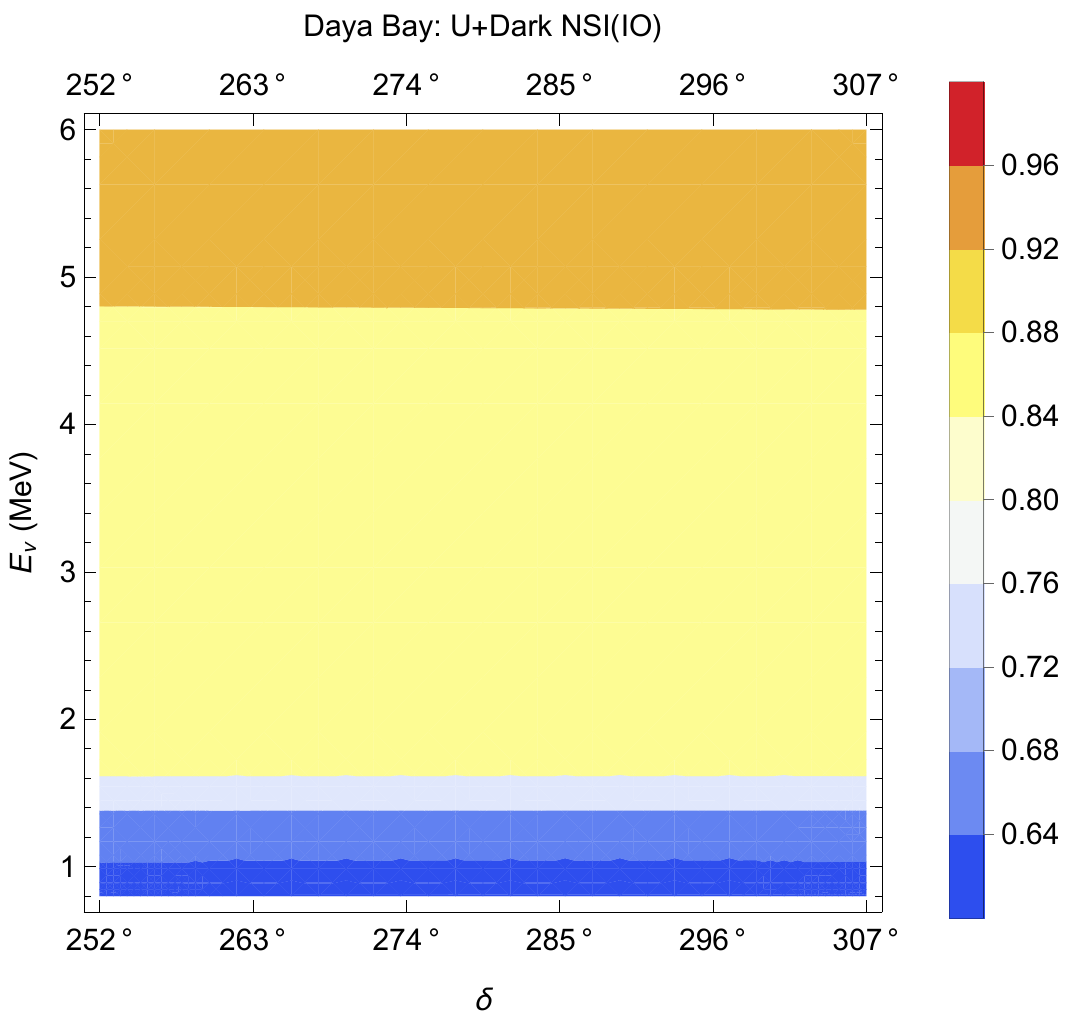}\hspace*{1mm} \includegraphics[scale=0.8]{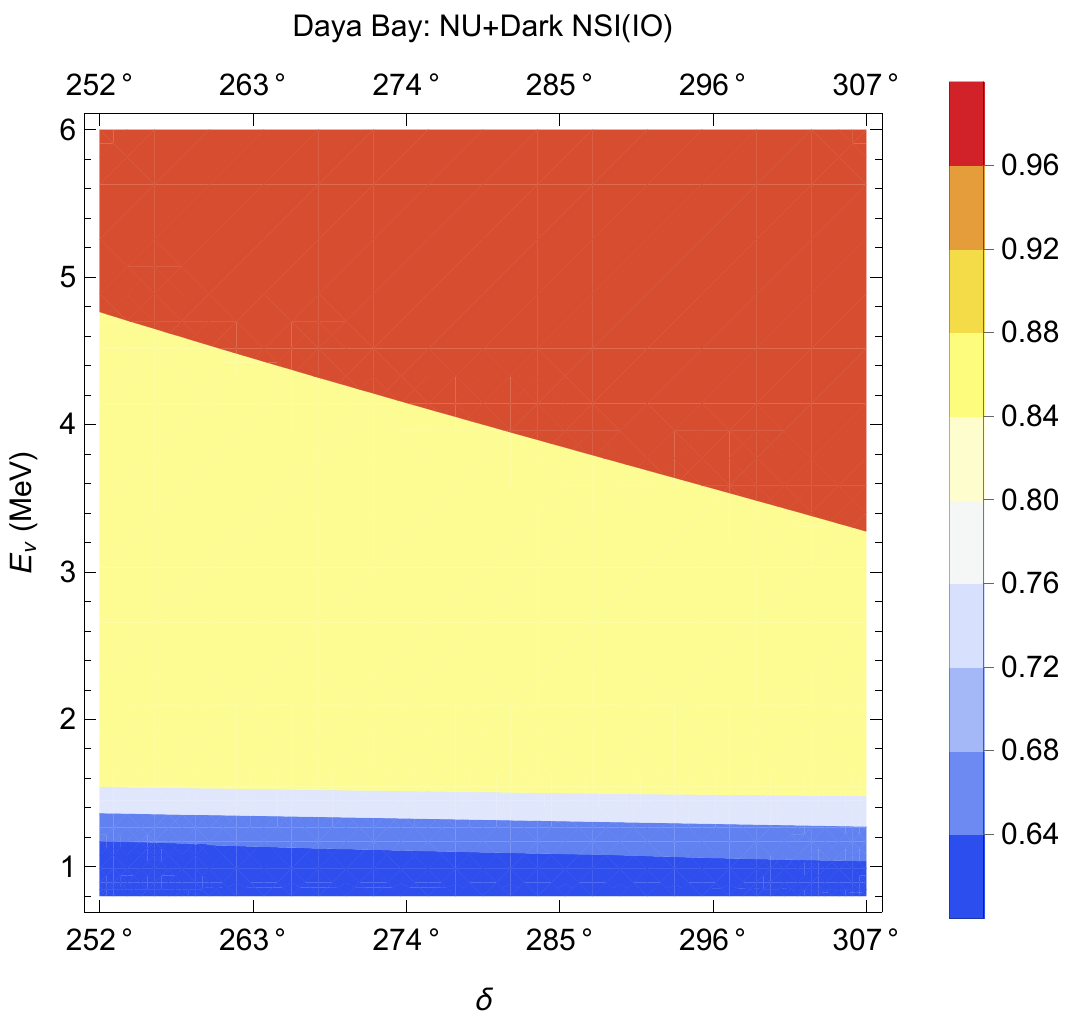}
    \caption{Same as fig. \ref{f7} for dark NSI.}
    \label{f4}
\end{figure*}

In fig. \ref{f5} and \ref{f6} similar results are presented for DUNE set up for IO. Here the maximum expanse of $P_{\mu e}$ is much lowered $i.e.~P_{\mu e, max}\gtrsim 0.09$, as compared to the case of NO. From the two figures, it is observed that $P_{\mu e}$ is very low in the major portion of the ($E_{\nu},\delta$) plane, for both of the NP scenarios of LV and dark NSI. Also, $P_{\mu e}$ is comparatively much lower in case of unitary mixing. For NU mixing in case of LV scenario, $P_{\mu e,max}$ is confined in a very tiny region at $(E_{\nu},\delta)\sim (1~\text{GeV}, 252^o)$, while for dark NSI, $P_{\mu e,max}$ corresponds to the range $252^o\lesssim \delta \lesssim 260^o$ around $E_{\nu}\sim1$ GeV. The results for fig. \ref{f1}, \ref{f2}, \ref{f5} and \ref{f6} are illustrated in table \ref{tt3}.

One distinguishable feature to be noticed from the above figures is that DUNE experiment is capable of exhibiting the signature of dark NSI in terms of large value of $P_{\mu e}(\gtrsim 0.225)$, when NO is taken into consideration, around the energy $E_{\nu}\approx 2.5-2.8$ GeV , which corresponds to the energy of the maximum flux.

Similar to DUNE, the estimation of the survival probability in $\bar{\nu}_e\rightarrow \bar{\nu}_e$ channel in context of Daya Bay experiment are presented in fig. \ref{f} and \ref{ff} , for LV and dark NSI scenario considering unitarity and non-unitarity of neutrino mixing. Here the survival probability of electron anti-neutrino ($\bar{P}_{ee}$) is plotted in ($E_{\nu}, \delta$) plane. 

\begin{center}
\begin{table}[h!]
\begin{tabular}{|c|c|c|c||c|c|} 
\hline
\multicolumn{6}{|c|}{Daya Bay} \\
\hline
 \multirow{2}{*}{\vtop{\hbox{\strut Mass}\hbox{\strut ordering}}} & ($E_{\nu}$,$\delta$) & \multicolumn{2}{c||}{Unitary (U)} & \multicolumn{2}{c|}{Non-unitary (NU)}\\
  \cline{3-6}
 & & LV & NSI &  LV & NSI  \\
  \hline
  \multirow{2}{*}{NO} & $E_{\nu}$ (MeV) &$1.8-2.3$ & - & \makecell{$1.8-2.8$,\\$1$} & $1.8-2.7$  \\
  \cline{2-6}
 & $\delta$ ($^o$) & $189-257$ & - & \makecell{$189-257$,\\$223-257$} & $189-257$ \\
  \hline
 \multirow{2}{*}{IO} & $E_{\nu}$ (MeV) & $1.7-2.3$ & - & \makecell{1.8-2.2,\\1} & $3.2-6$  \\
  \cline{2-6}
 & $\delta$ ($^o$) & $252-307$ & - & \makecell{252-307,\\265-307} & $252-307$ \\
  \hline
\end{tabular}
\caption{The specific regions of ($E_{\nu},\delta$) plane corresponding to $\bar{P}_{ee,max}\gtrsim 0.96$ (for both NO and IO), in the two NP scenarios of LV and dark NSI, under the assumption of unitary and non-unitary neutrino mixing, in the context of SBL Daya Bay experimental set up ($E_{\nu}=0.8-6$ MeV, $L=2$ km).}\label{t3}
\end{table}
\end{center}

In the scenario of LV, $\bar{P}_{ee}$ attains its maximum value ($\bar{P}_{ee,max}\gtrsim 0.96$) in a broad strip around $E_{\nu}\sim 2$ MeV within the entire allowed range of $\delta$ $i.e.$ $189^o\lesssim \delta \lesssim 257^o$ for both unitary and non-unitary mixing. Additionally, in the NU scenario, there is a narrow strip confined within $223^o \lesssim \delta \lesssim 257^o$ at $E_{\nu}\sim 1$ MeV. For dark NSI $\bar{P}_{ee,max}$ corresponds to a broad region around $E_{\nu}\approx 2$ MeV along the total range of $\delta$ for non-unitary mixing, while for unitary mixing $\bar{P}_{ee}$ does not even surpass $\sim 0.96$, as observed from the two panels of fig. \ref{ff}.

Similar plots are presented for IO in the two panels of fig. \ref{f7} and \ref{f4}. It is observed that the maximum value attained by $\bar{P}_{ee}$ remains the same as NO $i.e.$ $\bar{P}_{ee,max}\gtrsim 0.96$. In fig. \ref{f7}, $\bar{P}_{ee,max}$ corresponds to a  wide band spread along the full allowed range of $\delta$ $i.e.~252^o\lesssim \delta \lesssim 307^o$ for IO in case of unitary mixing, within $1.8~\text{MeV} \lesssim E_{\nu}\lesssim 2.3$ MeV. The region is slightly modified and becomes confined to a region within the energy range, $1.7~\text{MeV}\lesssim E_{\nu} \lesssim 2.1 $ MeV, for NU scenario. Moreover, $\bar{P}_{ee}$ also attains the large value ($\gtrsim 0.96$) around $E_{\nu}\approx 1$ MeV in a very thin strip , within $265^o \lesssim \delta \lesssim 307^o$.

In case of dark NSI, as shown in fig. \ref{f4}, $\bar{P}_{ee,max}$ corresponds to a very wide region in the energy interval $4.8~\text{MeV}\lesssim E_{\nu} \lesssim 6$ MeV at $\delta \sim 252^o$ for NU mixing which continues to enlarge further with increase in $\delta$. For unitary mixing $\bar{P}_{ee}$ remains well below $0.96$ in the entire $(E_{\nu},\delta)$ plane.

In case of Daya Bay, it is noticed that the large value of $\bar{P}_{ee}$ ($\gtrsim 0.96$) appears only in the scenario of dark NSI with NU neutrino mixing, in the energy range $E_{\nu}\approx 4-5$ MeV at which the excess $\bar{\nu}_e$ flux is observed than that predicted from the theory. The signature of LV is also apparent in Daya Bay set up for both NO and IO, at $E_{\nu}\approx 1$ MeV where $\bar{P}_{ee}\gtrsim 0.96$  and the energy resolution is quite good ($\sim 8.5\%$),  although it is more perceptible in case of NO.

Further, it can be pointed out from all the figures that the nature of $P_{\mu e}$ and $\bar{P}_{ee}$ in case of LV and dark NSI are quite different in case of both unitary and non-unitary neutrino mixing scenario in both the experimental set ups having different baseline lengths, which implies that these two BSM effects are completely non-degenerate, unlike the existent degeneracy between vector NSI and LV scenario \cite{sahoo2022core}. 

\section{Conclusion}\label{conclusion}
The effect of unitary and non-unitary mixing matrix on neutrino oscillation are analyzed and compared in the new physics scenarios of LV and dark NSI. The results are determined in the context of LBL and SBL experimental set up, DUNE and Daya Bay respectively, considering both kinds of neutrino mass orderings. In DUNE experiment the oscillation probability $P_{\mu e}$ is observed to be much higher for NO as compared to IO, in both the scenarios of LV and dark NSI. Also in case of NO, the incorporation of non-unitary mixing matrix is able to provide relatively higher value of $P_{\mu e}$ in certain regions of ($E_{\nu},\delta$) plane. An interesting feature observed here is that at $E_{\nu}\approx 2.5-2.8$ GeV large value  of $P_{\mu e}(\gtrsim 0.225)$ is noticed only in case of dark NSI for non-unitary mixing scenario. This points towards the existence of the new physics in the sector of dark NSI and shows the potential of DUNE to probe this sector 
at the energy corresponding to the maximum flux. 
However, no such distinguishable feature for LV is observed at DUNE.

The distinct signature of dark NSI is also observed in case of SBL Daya Bay set up. At   $E_{\nu}\approx 4-5$ MeV, $\bar{P}_{ee}$ possesses large value ($\gtrsim 0.96$) only in the scenario of dark NSI for IO, which could also be a possible explanation for the observed discrepancy in $\bar{\nu}_e$ flux at this energy interval. Further, $\bar{P}_{ee}$ is also large $(\gtrsim 0.96)$ at $E_{\nu}\approx 1$ MeV at which the energy resolution of the detector is  $\sim 8.5\%$ . This is only observed in the LV scenario for non-unitary neutrino mixing. Although such feature is noticed for both NO and IO, it is more prominently discernible in case of NO only.


\bibliographystyle{unsrt}
\bibliography{reference}

\end{document}